\documentclass[aps,prd,twocolumn,reprint,amsmath,amssymb,preprintnumbers,superscriptaddress,nobibnotes,nofootinbib,floatfix]{revtex4-1}
\usepackage{graphicx}
\usepackage{epstopdf}
\usepackage{amsfonts}
\usepackage{amssymb}
\usepackage{amsmath}
\usepackage{appendix}
\usepackage{enumerate}
\usepackage{natbib}
\usepackage{comment}
\usepackage{bbold}
\usepackage[shortlabels]{enumitem}
\usepackage{color}
\usepackage{slashed}
\usepackage{subfigure}
\usepackage{setspace}
\usepackage{footnote}
\usepackage{lipsum}
\usepackage{multirow}
\usepackage[colorlinks = true,
            linkcolor = blue,
            urlcolor  = blue,
            citecolor = blue,
            anchorcolor = blue, pdfusetitle]{hyperref}
\usepackage[capitalize]{cleveref}
\usepackage{braket}
\usepackage[compat=1.1.0]{tikz-feynman}
\usepackage{multirow}
\usepackage{physics}
\usepackage{feynmp-auto}
\usepackage[normalem]{ulem}
\usepackage{url}
\usepackage{units}
\usepackage[normalem]{ulem}
\usepackage{bm}

\newcommand{\be}{\begin{eqnarray}}
\newcommand{\ee}{\end{eqnarray}}
\newcommand{\ba} {\begin{equation}\begin{aligned}}
\newcommand{\ea} {\end{aligned}\end{equation}}
\newcommand{\bg} {\begin{equation}\begin{gathered}}
\newcommand{\eg} {\end{gathered}\end{equation}}

\newcommand{\beq}{\begin{equation}}
\newcommand{\eeq}{\end{equation}}

\usepackage{tikz,xcolor,hyperref}

\definecolor{lime}{HTML}{A6CE39}
\DeclareRobustCommand{\orcidicon}{\hspace{-1mm}
	\begin{tikzpicture}
		\draw[lime, fill=lime] (0,0) 
		circle [radius=0.12] 
		node[white] {{\fontfamily{qag}\selectfont \tiny \,ID}};
		\draw[white, fill=white] (-0.0525,0.095) 
		circle [radius=0.007];
	\end{tikzpicture}
	\hspace{-3mm}
}

\foreach \x in {A, ..., Z}{\expandafter\xdef\csname orcid\x\endcsname{\noexpand\href{https://orcid.org/\csname orcidauthor\x\endcsname}
		{\noexpand\orcidicon}}
}

\begin{document}
\setlength{\textfloatsep}{5pt}
\preprint{FERMILAB-PUB-25-0566-T, MITP-2025-045, CETUP-2025-004}

\title{Dark-matter-enhanced probe of relic neutrino clustering}
	\author{Writasree Maitra}
	\email{m.writasree@wustl.edu}
	\affiliation{Department of Physics, Washington University, St. Louis, Missouri 63130, USA}

\author{Anna M. Suliga}
	\email{a.suliga@nyu.edu}
	\affiliation{Center for Cosmology and Particle Physics, New York University, New York 10003, USA}

	\author{Vedran Brdar}
	\email{vedran.brdar@okstate.edu}
	\affiliation{Department of Physics, Oklahoma State University, Stillwater, Oklahoma, 74078, USA}
	
	\author{P. S. Bhupal Dev}
	\email{bdev@wustl.edu}
	\affiliation{Department of Physics, Washington University, St. Louis, Missouri 63130, USA}
	\affiliation{McDonnell Center for the Space Sciences, 
Washington University, St. Louis, MO 63130, USA}	
    \affiliation{PRISMA$^+$ Cluster of Excellence and Mainz Institute for Theoretical Physics, Johannes Gutenberg-Universit\"{a}t Mainz, 55099 Mainz, Germany}

\begin{abstract}
We propose heavy decaying dark matter (DM) as a new probe of the cosmic neutrino background (C$\nu$B).  Heavy DM, with mass $\gtrsim 10^9$ GeV,  decaying into neutrinos can be a new source of ultrahigh-energy (UHE) neutrinos. 
Including this contribution along with the measured astrophysical and predicted cosmogenic neutrino fluxes, we study the scattering of UHE neutrinos with the C$\nu$B via standard weak interactions mediated by the $Z$ boson. We solve the complete neutrino transport equation, taking into account both absorption and reinjection effects, to calculate the expected spectrum of UHE neutrino flux at future neutrino telescopes, such as  the IceCube-Gen2 radio.  We argue that such observations can be used to probe the C$\nu$B properties and, in particular, local C$\nu$B clustering. We find that, depending on the absolute neutrino mass and the DM mass and lifetime, a local C$\nu$B overdensity $\gtrsim 10^6$ can be probed at the IceCube-Gen2 radio within ten years of data taking. 
\end{abstract}

\maketitle

\section{Introduction}
\label{sec:Introduction}
Analogous to the cosmic microwave background (CMB), the standard cosmological model predicts the existence of a cosmic neutrino background (C$\nu$B) -- a relic population of neutrinos that decoupled from the primordial plasma approximately one second after the big bang, at temperatures on the order of 1 MeV~\cite{Weinberg:1962zza, Scott:2024rwc}. After decoupling, neutrinos free stream through the Universe, retaining a Fermi-Dirac distribution, with their temperature  evolution following an adiabatic expansion: $T_\nu\propto a^{-1}$, where $a(t)$ is the scale factor. After electron-positron annihilation, which heats the photon bath but not the neutrinos, the neutrino temperature becomes slightly lower than that of the photons. Consequently, the present-day C$\nu$B temperature can be written in terms of the CMB temperature as $ T_{\nu,0}=(4/11)^{1/3}T_{\gamma,0}\simeq 1.95~{\rm K}\simeq 1.68\times 10^{-4}~{\rm eV}$.
Assuming that the lightest neutrino mass $m_1\gtrsim T_{\nu,0}$, the C$\nu$B today can be thought of as a nonrelativistic gas of fermions with a number density per flavor and per degree of freedom
\begin{equation}
n_{\nu,0} =  \frac{3}{4}\frac{\zeta(3)}{\pi^2} T_{\nu,0}^3 \simeq 56~{\rm cm}^{-3} \ ,
\label{eq:nnu}
\end{equation}
and a total number density of $6\,n_{\nu,0}$ (valid irrespective of whether neutrinos are Dirac or Majorana~\cite{Long:2014zva,Bauer:2022lri}), considering three neutrino flavors and both neutrinos and  antineutrinos.

The presence of C$\nu$B has only been indirectly confirmed through its imprints on (i) the big bang nucleosynthesis via the expansion rate and neutron-to-proton ratio~\cite{Steigman:2012ve}, (ii) the anisotropies in the CMB power spectrum~\cite{Planck:2018vyg}, (iii) the formation and distribution of large-scale cosmological structures~\cite{DESI:2025ejh}, and (iv) acoustic oscillation phase shifts~\cite{Follin:2015hya}. Despite several proposed strategies~\cite{Stodolsky:1974aq, Weinberg:1962zza, Irvine:1983nr, Gelmini:2004hg, 10.3389/fphy.2014.00030, PTOLEMY:2018jst, PTOLEMY:2022ldz, Long:2014zva, Bauer:2022lri, delCastillo:2025qnr}, the direct detection of C$\nu$B remains elusive.  
The extremely low kinetic energy of C$\nu$B and the weakly interacting nature of neutrinos are the biggest obstacles to observing these relic neutrinos directly in the terrestrial laboratories via their scattering. 

Therefore, an interesting alternative is to flip the script and consider the impact of the C$\nu$B as scatterers instead of scattering particles. The first such proposal is the so-called $Z$ burst~\cite{Weiler:1982qy} (see also Refs.~\cite{Roulet:1992pz,Yoshida:1996ie,Fargion:1997ft, Weiler:1997sh,Yoshida:1998it, Eberle:2004ua, Barenboim:2004di}), where ultrahigh-energy (UHE) neutrinos scatter off  C$\nu$B through a resonant process in which an on-shell $Z$ boson is exchanged.  
Such a process could, in principle, leave a distinct absorption feature in the cosmogenic neutrino flux measured at Earth. However, the neutrino energy required to produce an on-shell $Z$ boson, $E_\nu = m_Z^2/2m_\nu(1+z)\gtrsim 10^{13}$ GeV ($z$ being the redshift at which the $\nu\bar{\nu}$ annihilation occurs in the sky), far exceeds the Greisen–Zatsepin–Kuzmin (GZK) cutoff~\cite{Greisen:1966jv, Zatsepin:1966jv}, which sets a theoretical upper bound on the energy of cosmic rays that are the progenitors of the cosmogenic neutrinos~\cite{Berezinsky:1969erk}. The resonance energy gets smaller for the on-shell production of less massive particles compared to the $Z$ boson. This was employed in Ref.~\cite{Brdar:2022kpu} for the case of $\rho$ meson exchange within the Standard Model (SM), where the resonance occurs at $E_\nu = m_\rho^2/2m_\nu(1+z)\gtrsim 10^{9}$ GeV. By numerical coincidence, the position of the $\rho$ resonance overlaps with the peak of the predicted cosmogenic neutrino flux, which offers an enhanced absorption effect at higher redshifts. In presence of new (light) scalars or gauge bosons, similar resonances might occur, even at lower energies~\cite{Ng:2014pca, Ibe:2014pja,Araki:2014ona,Kamada:2015era,DiFranzo:2015qea,Altmannshofer:2016brv}. However, there are two major challenges in searching for  resonance-induced absorption features in the cosmogenic neutrino spectrum: (i) The width of the resonance must be comparable to or greater than the resolution of the experiment in order to resolve the dip feature. This happens to be the case for the $\rho$ meson which is the lightest vector meson in the SM and has a width-to-mass ratio of 19\%~\cite{ParticleDataGroup:2024cfk}. In contrast, almost all other SM resonances have narrow widths~\cite{ParticleDataGroup:2024cfk, Dev:2021tlo}, thus making the $\rho$ meson possibly the only SM resonance in the $\nu\bar{\nu}$ channel relevant at future neutrino telescopes. (ii) There is a large uncertainty on the cosmogenic neutrino flux predictions~\cite{Berezinsky:1969erk, Engel:2001hd, Hooper:2004jc, Allard:2006mv, Ahlers:2010fw, Kotera:2010yn, Kampert:2012mx, Ahlers:2012rz, Aloisio:2015ega, AlvesBatista:2018zui, Moller:2018isk, vanVliet:2019nse, Heinze:2019jou, Muzio:2019leu, Muzio:2021zud, Valera:2022wmu, Ehlert:2023btz, Muzio:2023skc}, which is mainly due to the current uncertainty in the cosmic ray composition at the highest energies and the unknown cosmic ray source evolution history. The recent measurements from Auger~\cite{PierreAuger:2019ens} and IceCube~\cite{IceCubeCollaborationSS:2025jbi} have already excluded some of the cosmogenic flux models mentioned above;  however, there still remains almost an order of magnitude uncertainty in the flux prediction among the allowed models, thus making it further difficult to search for a dip feature.   

In light of these difficulties, we complement Ref.~\cite{Brdar:2022kpu} in the present work in two aspects. First, instead of resonant production, we focus on off-shell $Z$ boson-induced scattering of UHE neutrinos with C$\nu$B~\cite{Roulet:1992pz} and hence consider a wide energy range rather than resonance energies only. Second, in addition to astrophysical and cosmogenic neutrinos, we add the neutrino flux contribution from the decay of superheavy dark matter (SHDM) particles~\cite{Gelmini:1999ds}. Specifically, we consider decaying neutrinophilic dark matter (DM) particles with mass $m_{\text{DM}}\gtrsim 10^{9}$ GeV. Such an alternative DM scenario is well motivated~\cite{Chung:1998ua}, especially in the era of the waning of the weakly interacting massive particle (WIMP) paradigm~\cite{Arcadi:2024ukq}. SHDM easily evades the direct detection and collider constraints, unless they are strongly interacting~\cite{Albuquerque:2003ei}. Moreover, the standard indirect detection signals from DM annihilation are highly suppressed, because of the small annihilation cross section $\sigma v$ and small DM number density $n_{\rm DM}$. The indirect detection prospects for SHDM improve if the  DM is not absolutely stable as the decay signal scales like $n_{\rm DM}$ (instead of $n_{\rm DM}^2$ as in annihilation). This is conceivable because discrete symmetries typically invoked to stabilize DM are expected to be broken at high-energy scales by quantum gravity effects. Note that the SHDM production in the early Universe usually requires a nonthermal mechanism to obtain the correct relic abundance, since the thermal freeze-out due to DM annihilation in a standard cosmological history results in DM overproduction for masses larger than $\sim 100$ TeV--the so-called unitarity bound~\cite{Griest:1989wd}.\footnote{For possible ways to relax or evade the unitarity bound, see, e.g., Ref.~\cite{Nussinov:2025dkc}.} Explicit constructions for SHDM production and decay have been considered in inflationary models~\cite{Chung:1998zb, Chung:2001cb,Kannike:2016jfs,Kolb:2017jvz,Li:2019ves,Haro:2019umj,Babichev:2020yeo,Ling:2021zlj,Verner:2024agh,Murase:2025uwv}, in string theory~\cite{Benakli:1998ut,Allahverdi:2020uax,Allahverdi:2023nov, Ling:2025nlw}, and in seesaw framework~\cite{Uehara:2001wd,Huong:2016ybt,Dev:2016qbd,Sui:2018bbh,Deligny:2024fyx}.  Here we take a model-independent, phenomenological approach and simply treat the SHDM mass $m_{\rm DM}$ and lifetime $\tau_{\rm DM}$ as free parameters. We will choose representative benchmark points such that  these quantities are consistent with the existing limits, as summarized in \cref{sec:DMtau}. We assume that the correct DM relic abundance can always be obtained for these benchmarks, depending on  the details of the explicit model construction and cosmological history; see \cref{sec:DMprod} for a short discussion and parametric estimate indicating the relevant cosmological scales. 

Including the two above-mentioned effects, namely, (i) the off-shell $Z$ contributions and (ii) the DM-induced neutrino flux, we find that future neutrino telescopes like IceCube-Gen 2 radio~\cite{IceCube-Gen2:2020qha} can probe the C$\nu$B, but only in the presence of a local overdensity. We will denote by $\xi$ the dimensionless ratio of the testable C$\nu$B number density to the standard cosmological value given in Eq.~\eqref{eq:nnu}. 

As neutrinos are massive, the C$\nu$B can become clustered because of the gravitational potential felt from both baryonic matter and DM. Hence, the local neutrino density is expected to be enhanced; however, such an enhancement can at most be at the level of few tens of percent in the Milky Way~\cite{Ringwald:2004np, deSalas:2017wtt, Mertsch:2019qjv, Hotinli:2023scz, Zimmer:2023jbb, Holm:2024zpr}. Nevertheless, very large overdensities of C$\nu$B can be achieved in the presence of new physics. For example, attractive Yukawa interactions of neutrinos with a new light scalar boson can lead to the formation of stable bound states and neutrino clusters~\cite{Wise:2014ola, Smirnov:2022sfo} with maximal central density up to $10^9~{\rm cm}^{-3}$, which translates to an overdensity parameter $\xi\sim 10^7$.

The current most stringent local constraint on $\xi$ comes from the Karlsruhe tritium neutrino (KATRIN) experiment: $\xi < 1.1 \times 10^{11}$ at 95\% C.L.~\cite{KATRIN:2022kkv}. The high-precision tracking of astronomical objects in Solar System places comparable constraints of the same order~\cite{Tsai:2022jnv}. These two probes are sensitive to overdensities located closer than 10 A.U. from Earth. For other astrophysical probes including overdensities across larger distances, see Refs.~\cite{Bauer:2020jay, Brdar:2022kpu,
Ciscar-Monsalvatje:2024tvm, Franklin:2024amy, DeMarchi:2024zer, Herrera:2024upj, Dev:2024yrg, Chauhan:2024deu, Das:2024thc, Zhang:2025rqh}, where the derived constraints vary between $\xi \sim 10^6$ and $\xi \sim 10^{14}$. The strongest astrophysical bound comes from relic neutrino  upscattering with ultrahigh-energy cosmic rays (UHECRs)~\cite{Herrera:2024upj, Zhang:2025rqh}, but this crucially depends on the UHECR composition and source evolution, both of which are largely uncertain at the moment.  In this work, by comparing expected neutrino event rates induced by heavy DM decay with and without overdensity at the future IceCube-Gen2 radio experiment~\cite{IceCube-Gen2:2020qha, IceCube-Gen2:2021rkf, Glaser:2025ler}, we derive a projected sensitivity of $\xi \gtrsim 10^6$, assuming a single overdense C$\nu$B cloud centered at redshift $z = 0$. The radius of the cloud $R$ is calculated assuming that the total number of relic neutrinos matches the one in standard cosmology and is given by $R = \frac{c}{H_0} \xi^{-1/3}$, where $H_0$ is the Hubble constant today and $c$ is the speed of light in vacuum. Our projected sensitivity is comparable to those from other astrophysical probes and is significantly better than the current laboratory constraint. 

The rest of this paper is organized as follows. In \cref{sec:flux}, we discuss the production of UHE neutrinos from the decay of heavy DM particles and from astrophysical environments, as well as their evolution to Earth. In \cref{sec:constraints}, we use these UHE neutrino fluxes to obtain IceCube-Gen2 radio event rates and we derive the corresponding sensitivity to a C$\nu$B overdensity. We conclude in \cref{sec:Conclusion}. In Appendix~\ref{sec:DMtau}, we summarize the existing constraints on SHDM mass and lifetime. In Appendix~\ref{sec:transport}, we present the neutrino transport equation.

\section{Ultrahigh-Energy Neutrino Flux}
\label{sec:flux}

We consider a heavy scalar DM, with masses above $10^9$ GeV, that can decay into neutrino-antineutrino pairs of all three flavors ($\text{DM}\rightarrow \nu_\alpha \bar{\nu}_\alpha$). The differential flux of neutrinos and antineutrinos produced in this way has two components: galactic and extragalactic. Extragalactic neutrinos encounter a significantly larger fraction of the C$\nu$B cloud on their way to Earth, and hence we focus exclusively on this contribution. Specifically, including the galactic component would worsen the sensitivity to C$\nu$B, simply because this component of the flux is hardly impacted through scattering with C$\nu$B given the much shorter propagation distances involved. 
Since radio experiments such as IceCube-Gen2 radio are expected to have exceptional angular resolution  below a few degrees~\cite{Hallgren2021Gen2Radio,Bouma:2023koi}, neutrinos coming from the direction of the Galactic Center can be isolated, as recently demonstrated by IceCube with the detection of neutrinos from the Galactic Plane~\cite{IceCube:2023ame}.

\begin{figure}[t]
\centering
\includegraphics[width=0.95\linewidth]{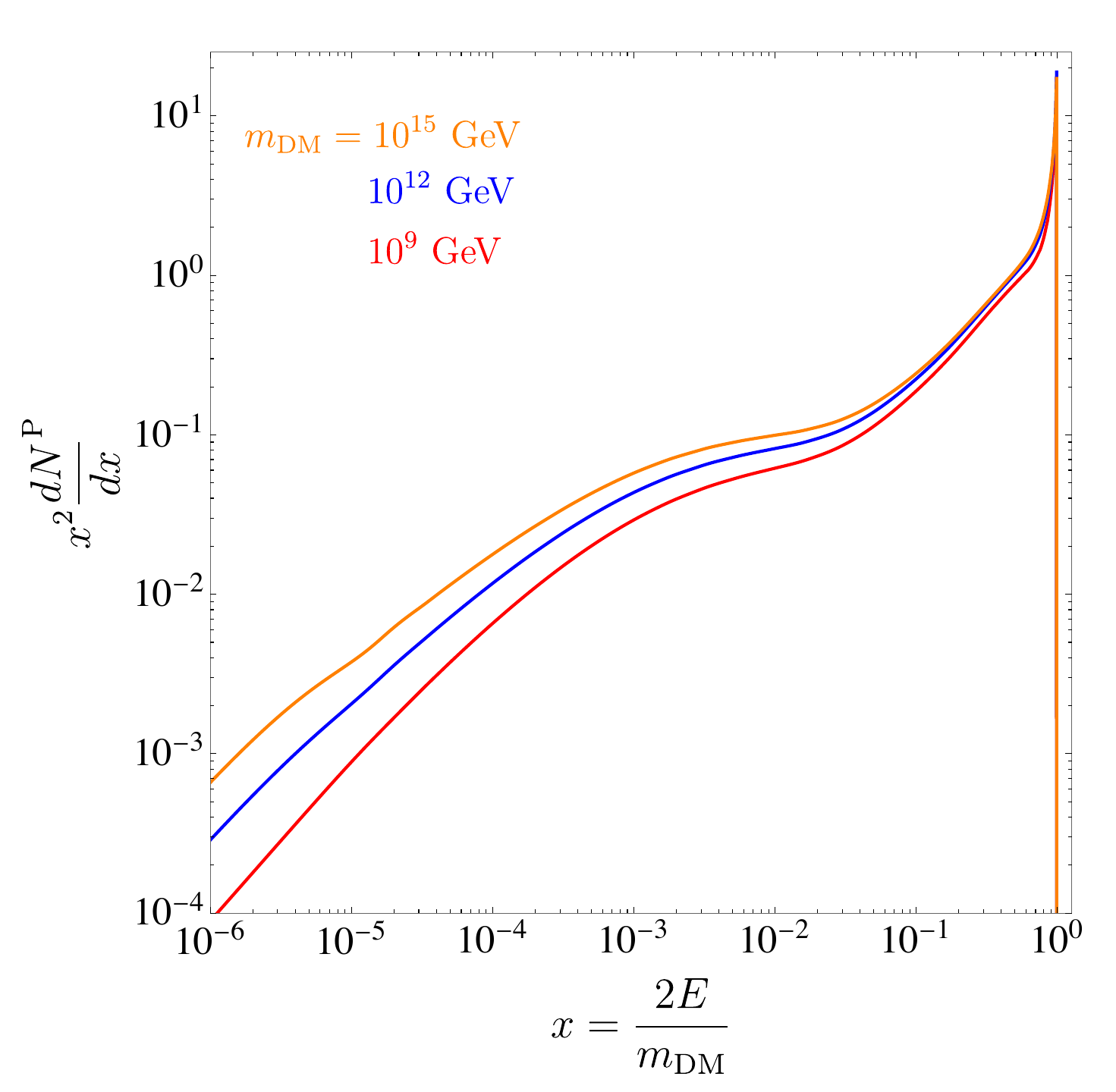}
\caption{
The prompt energy spectra of (anti)neutrinos resulting from DM decay into neutrino-antineutrino pairs summed over all flavors. We present results for three different DM masses: $m_\mathrm{DM}=10^9$ (red),  $10^{12}$ (blue), and  $10^{15}\;\mathrm{GeV}$ (orange). The spectra are generated using \texttt{HDMSpectra}~\cite{Bauer:2020jay}.}
\label{fig:DMspec}
\end{figure}

The differential flux of extragalactic neutrinos and antineutrinos of flavor $\alpha$ reads~\cite{Esmaili:2012us, Brdar:2016thq}
\begin{equation}
    \frac{d\Phi_{\nu_\alpha+\bar{\nu}_\alpha}^{\text{DM}}}{dE}=\frac{\Omega_\mathrm{DM}\rho_c}{4\pi m_\mathrm{DM}\tau_\mathrm{DM}}\int_0^\infty \frac{dz}{H(z)}\frac{dN_\alpha}{dE}(E,z)\,. 
    \label{eqnEG}
\end{equation}
Here, $\rho_c$ is the critical energy density of the Universe, $\Omega_\mathrm{DM}$ is the relic abundance of DM, $z$ is the cosmological redshift and $H(z)=H_0\sqrt{\Omega_\Lambda+\Omega_{\rm m}(1+z)^3}$ is the Hubble expansion rate with the dark energy and matter density components $\Omega_\Lambda=0.685$, $\Omega_{\rm m}=0.315$, and the Hubble constant $H_0=100\, h$ km.s$^{-1}$.Mpc$^{-1}$ with $h=0.673$~\cite{Planck:2018vyg}. The quantity $dN_\alpha/dE$ in the absence of neutrino interactions with C$\nu$B would simply be equal to the prompt energy spectrum $dN_\alpha^{\rm P}/dE$ (appropriately redshifted to account for cosmic expansion) that we calculate using \texttt{HDMSpectra}~\cite{Bauer:2020jay} and shown in \cref{fig:DMspec}.

However, UHE neutrinos produced from DM decays can scatter while propagating in the C$\nu$B, and this is what is captured by the neutrino transport equation~\cite{Ng:2014pca, Bustamante:2020mep, Creque-Sarbinowski:2020qhz, Esteban:2021tub, Das:2024bed} (see Appendix~\ref{sec:transport}). Specifically, this equation accounts for energy redshift due to the expansion of the Universe, as well as neutrino absorption and reinjection effects from scattering on the C$\nu$B.
We adopt the cross sections for $\nu$-$\nu$, $\bar{\nu}$-$\bar{\nu}$, and $\nu$-$\bar{\nu}$  interactions through $Z$ boson exchange from Ref.~\cite{Roulet:1992pz} (see also Ref.~\cite{Dev:2021tlo}). For brevity, in what follows we will just refer to all these processes simply by $\nu$-$\nu$. While we have investigated all the aforementioned terms in the transport equation, to keep the numerical treatment efficient and the runtime manageable, we eventually solved the transport equation by ignoring the reinjection terms, which we found not to significantly impact our final results. This is consistent with what was shown in Ref.~\cite{Das:2024bed}. With this procedure, we calculate the neutrino flux at Earth from extragalactic DM decays.

We also consider UHE neutrinos produced in the astrophysical environments through baryonic interactions. This additional flux consists of two types--cosmogenic and astrophysical neutrinos. The cosmogenic neutrinos are created when the cosmic ray protons or nuclei, during their propagation, interact with the CMB~\cite{Berezinsky:1969erk, Engel:2001hd, Hooper:2004jc, Allard:2006mv, Ahlers:2010fw, Kotera:2010yn, Kampert:2012mx, Ahlers:2012rz, Aloisio:2015ega, AlvesBatista:2018zui, Moller:2018isk, vanVliet:2019nse, Heinze:2019jou, Muzio:2019leu, Muzio:2021zud, Valera:2022wmu, Ehlert:2023btz, Muzio:2023skc}.
The astrophysical neutrinos, on the other hand, are produced when the cosmic ray protons or heavier nuclei that are generated at the cosmic reservoirs or accelerators undergo interactions with each other or with the ambient photons, and produce neutrinos as by-products~\cite{Meszaros:2017fcs}. Similarly as for DM flux, we employ the transport equation in order to propagate cosmogenic and astrophysical neutrinos through the C$\nu$B cloud.

\begin{figure}[t]
\centering
\includegraphics[width=0.99\linewidth, trim={0 1.55cm 0 0},clip]
{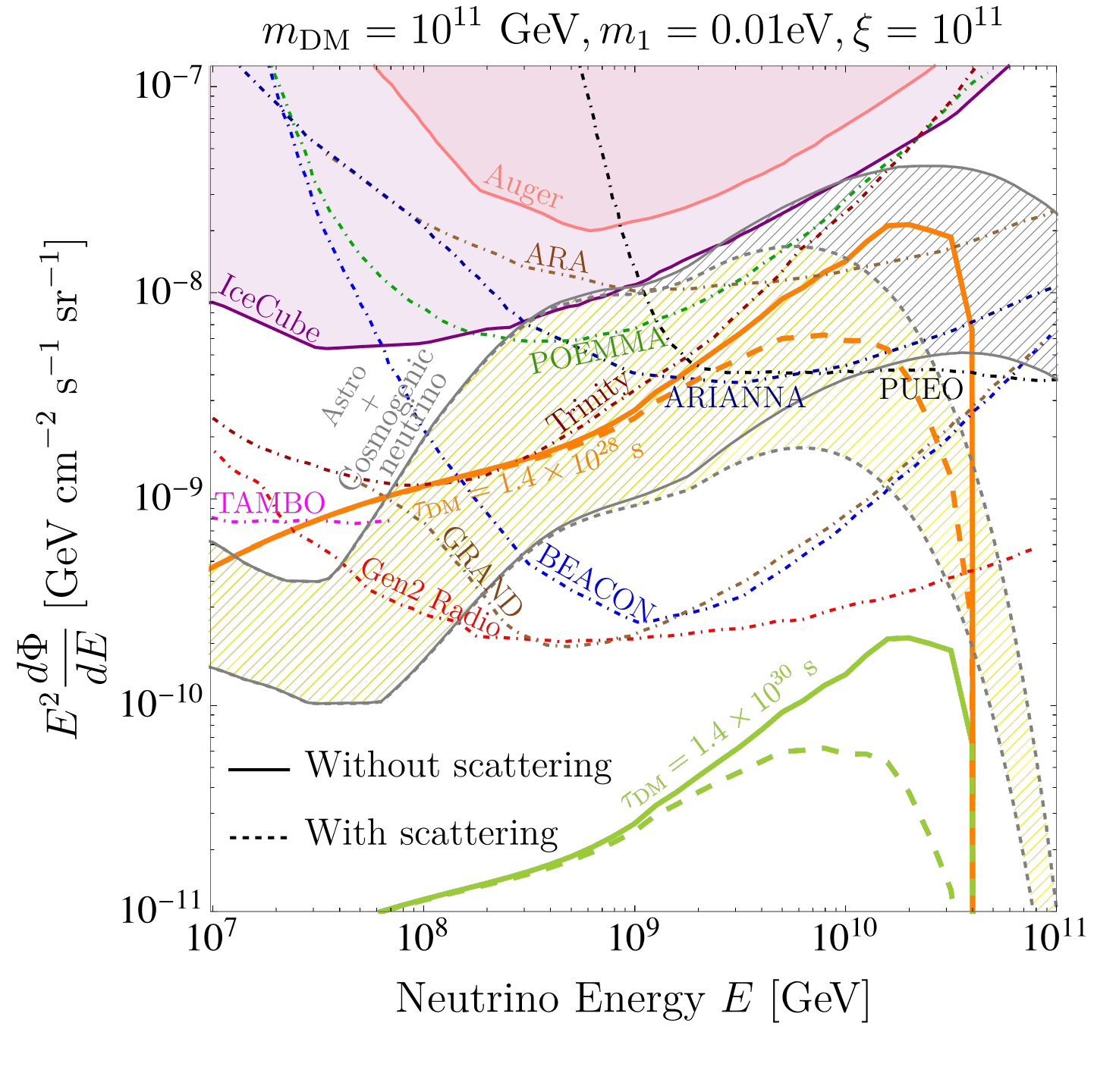} \hspace{0.5mm}
\caption{Comparison of neutrino fluxes, summed over all flavors, at Earth without (solid) and with (dashed) scattering off the C$\nu$B cloud with overdensity $\xi=10^{11}$ for normal neutrino mass ordering with lightest neutrino mass fixed at $0.01$ eV. For DM-induced flux component, we show two benchmark points for $m_\text{DM}=10^{11}$ GeV,  one that saturates the lifetime constraint from $\gamma$ rays (green) and another one that saturates the constraint from neutrinos  (orange) --see~\cref{fig:tauConstraints}. The gray hatched band illustrates the uncertainty in the prediction of the cosmogenic neutrino fluxes (taken from Ref.~\cite{Muzio:2023skc}), whereas the astrophysical flux was taken from Ref.~\cite{IceCube:2025ewu}. The light yellow band shows the astrophysical plus cosmogenic flux after scattering with the C$\nu$B cloud. The shaded regions in the upper half-plane are the current exclusion limits from Auger~\cite{PierreAuger:2019ens} and IceCube~\cite{IceCubeCollaborationSS:2025jbi}, while the other dot-dashed curves are the sensitivities of future experiments (taken from Ref.~\cite{Ackermann:2022rqc}).}
\label{DM-bkg-flux}
\end{figure}

In~\cref{DM-bkg-flux}, we show both DM-induced and astrophysical plus cosmogenic neutrino fluxes at Earth, comparing scenarios with and without scattering off an overdense C$\nu$B cloud. We take $\xi=10^{11}$ (which saturates the current KATRIN bound~\cite{KATRIN:2022kkv}) for illustration and assume normal neutrino mass ordering with the lightest neutrino mass $m_1=0.01$ eV. For the DM-induced flux component, we show two benchmark points with $m_\text{DM}=10^{11}$ GeV, one that saturates the lifetime constraint from $\gamma$ rays (green) and another one that saturates the lifetime constraint from neutrinos (orange) --see~\cref{fig:tauConstraints}. The cosmogenic neutrino flux with the associated uncertainties (as shown by the gray hatched band) is taken from Ref.~\cite{Muzio:2023skc}, derived by fitting the observed cosmic ray spectrum and composition data from Pierre Auger~\cite{PierreAuger:2020qqz}. Here, the upper and lower lines correspond to the active galactic nucleus (AGN) and star formation rate (SFR) source evolution models, respectively.   For the astrophysical flux component, we use a recent IceCube best-fit spectrum using a broken power law ~\cite{IceCube:2025ewu}.  The light yellow band shows the astrophysical plus cosmogenic flux after scattering with the C$\nu$B cloud. 
Clearly, the scattering in regions with C$\nu$B overdensity alters the predicted spectrum, and this is precisely what our analysis is based on. 
In~\cref{DM-bkg-flux}, we also show the flux upper limits from Auger~\cite{PierreAuger:2019ens} and IceCube~\cite{IceCubeCollaborationSS:2025jbi} (shaded regions in the upper half-plane), as well as the sensitivities of a number of future neutrino telescopes (taken from Ref.~\cite{Ackermann:2022rqc}), including IceCube-Gen2 radio~\cite{IceCube-Gen2:2020qha, IceCube-Gen2:2021rkf} which we will use as a representative example in our analysis.

\begin{figure*}[t!]
\centering
\includegraphics[height=0.49\linewidth,width=0.49\linewidth]{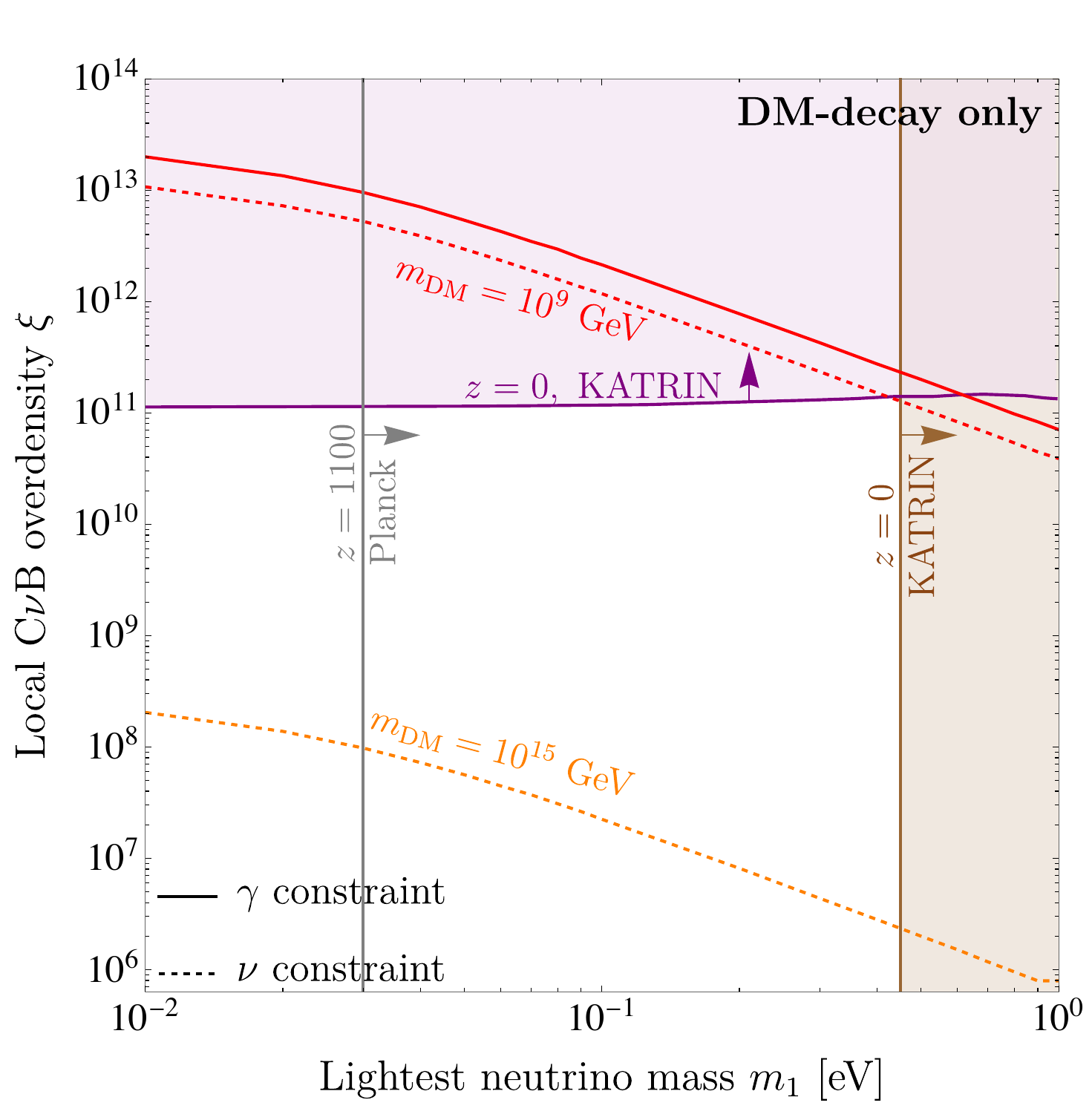}
\includegraphics[height=0.49\linewidth,width=0.49\linewidth]{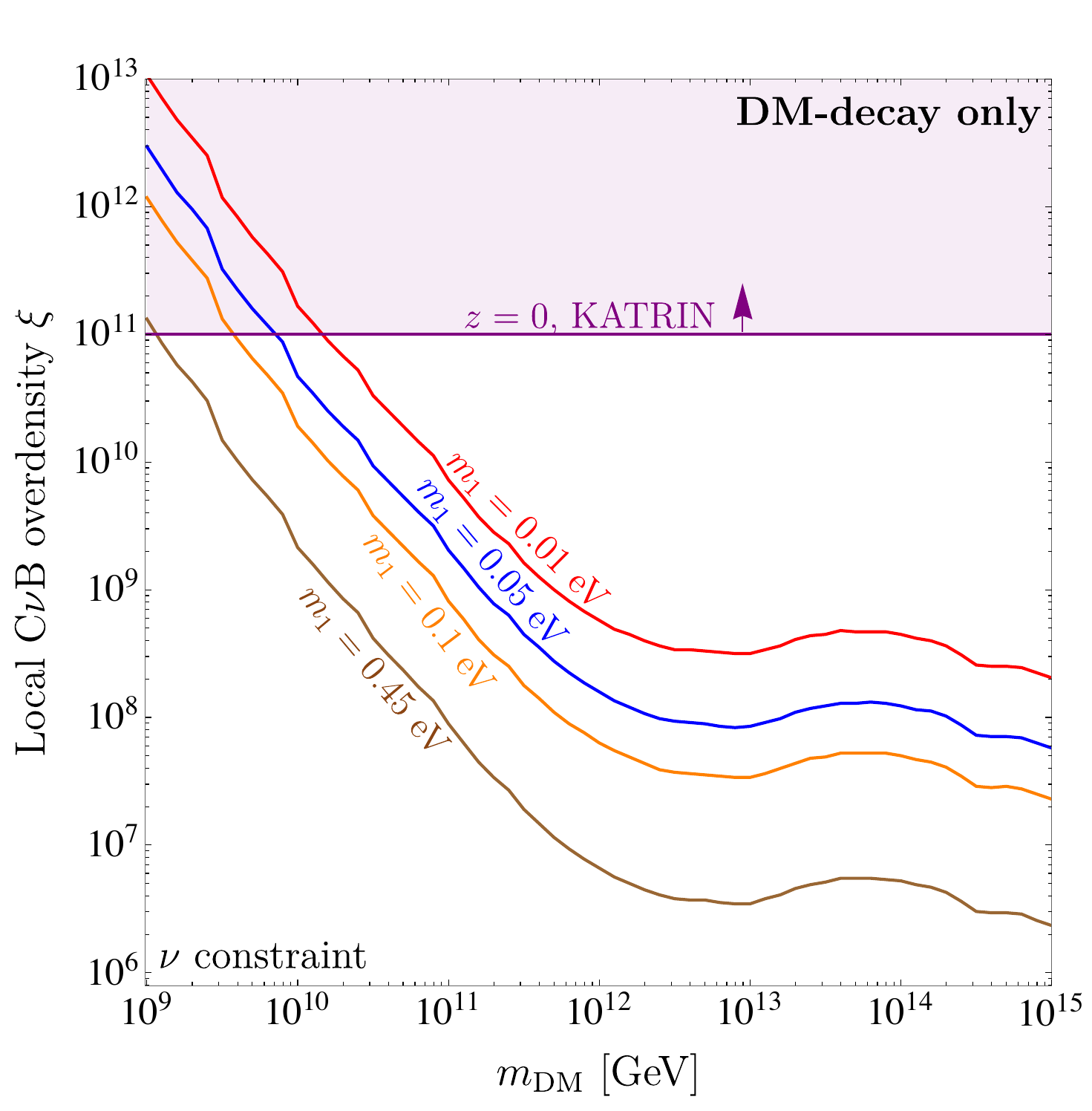}
\caption{\emph{Left:} projected IceCube-Gen2 radio sensitivity at 90\% C.L. on the C$\nu$B overdensity $\xi$ as a function of the lightest neutrino mass for two specific values of DM mass. Here we have considered the neutrino flux from DM decay only. Solid (dashed) lines correspond to cases where the DM-induced neutrino flux is computed for the DM parameters saturating the $\gamma$-ray (neutrino) constraints on DM decay. The horizontal purple line is the current 95\% C.L. KATRIN upper limit on the local C$\nu$B overdensity~\cite{KATRIN:2022kkv}, and the vertical brown line is the 90\% C.L. KATRIN upper limit on the absolute neutrino mass~\cite{KATRIN:2024cdt}.
The vertical gray line represents the cosmological upper limit from Planck~\cite{Planck:2018vyg}. 
\emph{Right:} projected IceCube-Gen2 radio sensitivity at 90\% C.L. on the C$\nu$B overdensity $\xi$ as a function of the DM mass for four specific values of the lightest neutrino mass. Here the DM-induced neutrino flux corresponds to the DM parameters saturating the neutrino constraint.   
}
\label{fig:dm_only_m1}
\end{figure*}
\section{Event Rates and Sensitivity to C$\nu$B Overdensity}
\label{sec:constraints}

The number of predicted neutrino events in IceCube-Gen2 radio experiment can be written as 
\begin{align}
N_{\text{w} (\text{w/o})} = T\int_{E_\nu^\mathrm{min}}^{E_\nu^\mathrm{max}} dE\int d\Omega ~ A_\mathrm{eff}(E,\Omega) \left(\frac{d\Phi_{\nu_\alpha+\bar{\nu}_\alpha}}{dE}\right)_{\text{w} (\text{w/o})},
\label{eq:N}
\end{align}
where w (w/o) in the subscript denotes with (without) including scattering off the overdense neutrino cloud. Further, in \cref{eq:N}, $T=10\;\mathrm{yr}$ is the assumed data taking time and $A_\mathrm{eff}(E,\Omega)$ is the effective area of the IceCube-Gen2 radio detector which is a function of the neutrino energy and solid angle. Given the available solid-angle average effective area of IceCube-Gen2, we have $\int d\Omega A_{\textrm{eff}}(E,\Omega)=4\pi A_{\textrm{eff}}(E)$. The energy integration limits in \cref{eq:N} are taken to be  $E_{\nu}^\mathrm{min} = 10^{7.5}\;$ and $E_{\nu}^\mathrm{max}=10^{15}\;$GeV. The IceCube-Gen2 radio effective area is available for the energy range between $10^{7.5}\;$GeV and $10^{11}\;$GeV~\cite{IceCube-Gen2:2021rkf}; above $10^{11}\;$GeV, we conservatively assume the effective area to be constant, although a $\log(E)$ growth is expected.

We calculate the projected sensitivity on the local C$\nu$B overdensity, $\xi$, assuming that the likelihood ratio test statistic follows the $\chi$-squared distribution with 1 degree of freedom, 
\begin{equation}
\label{eq:chisquare}
\chi^2 = 2\left( N_\mathrm{w} - N_\mathrm{w/o} + N_\mathrm{w/o} \log \frac{N_\mathrm{w/o}}{\mathrm{N_\mathrm{w}}} \right) \ .
\end{equation}
For the IceCube-Gen2 radio, to detect the signature of $\nu$-$\nu$ scattering at 90\% C.L., the value of $\chi^2$ needs to exceed $2.7$. By utilizing this condition, we calculate the projected sensitivity on $\xi$.

\begin{figure*}[t!]
\centering
\includegraphics[height=0.488\linewidth,width=0.49\linewidth]{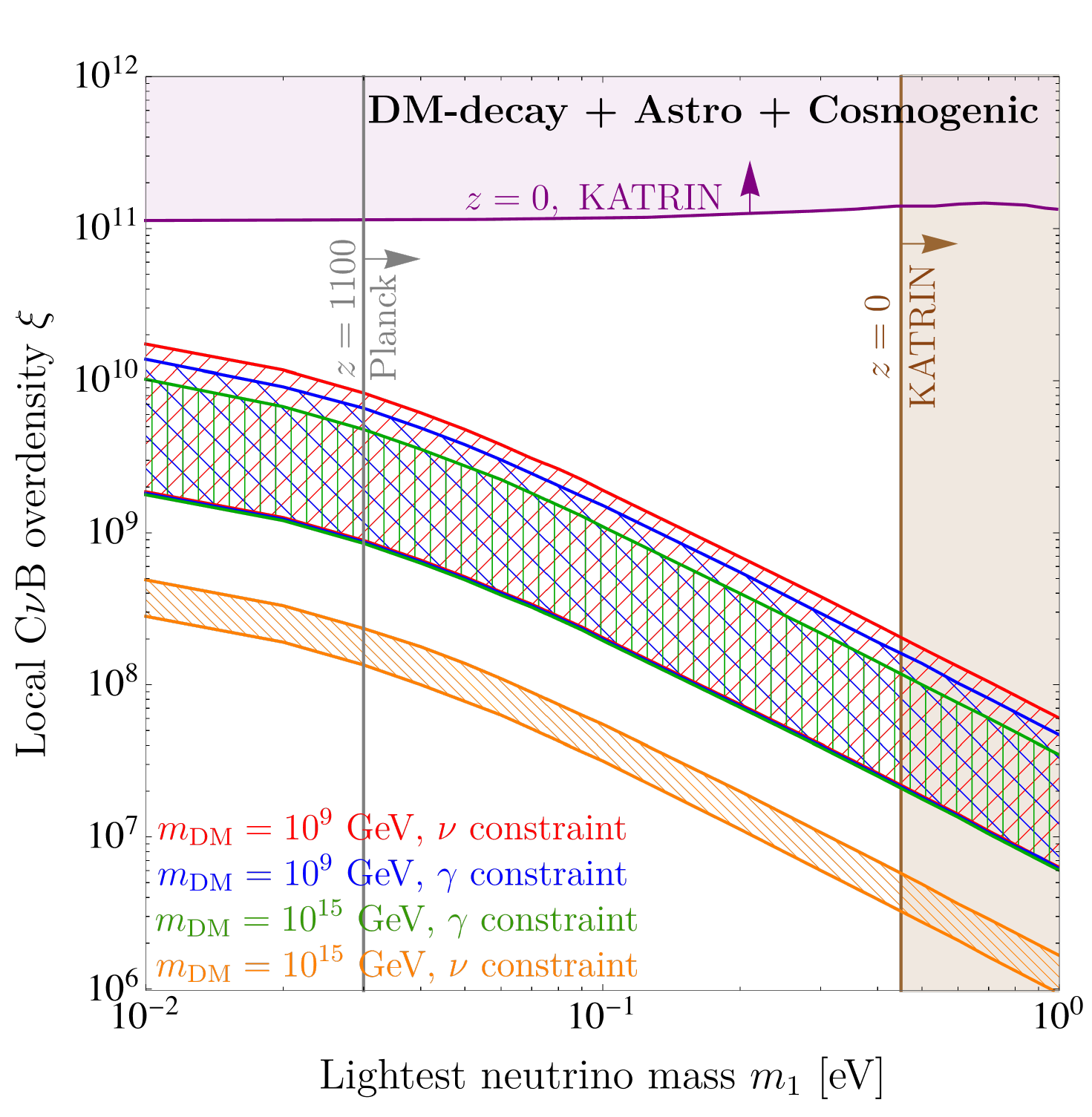}
\includegraphics[height=0.49\linewidth,width=0.49\linewidth]{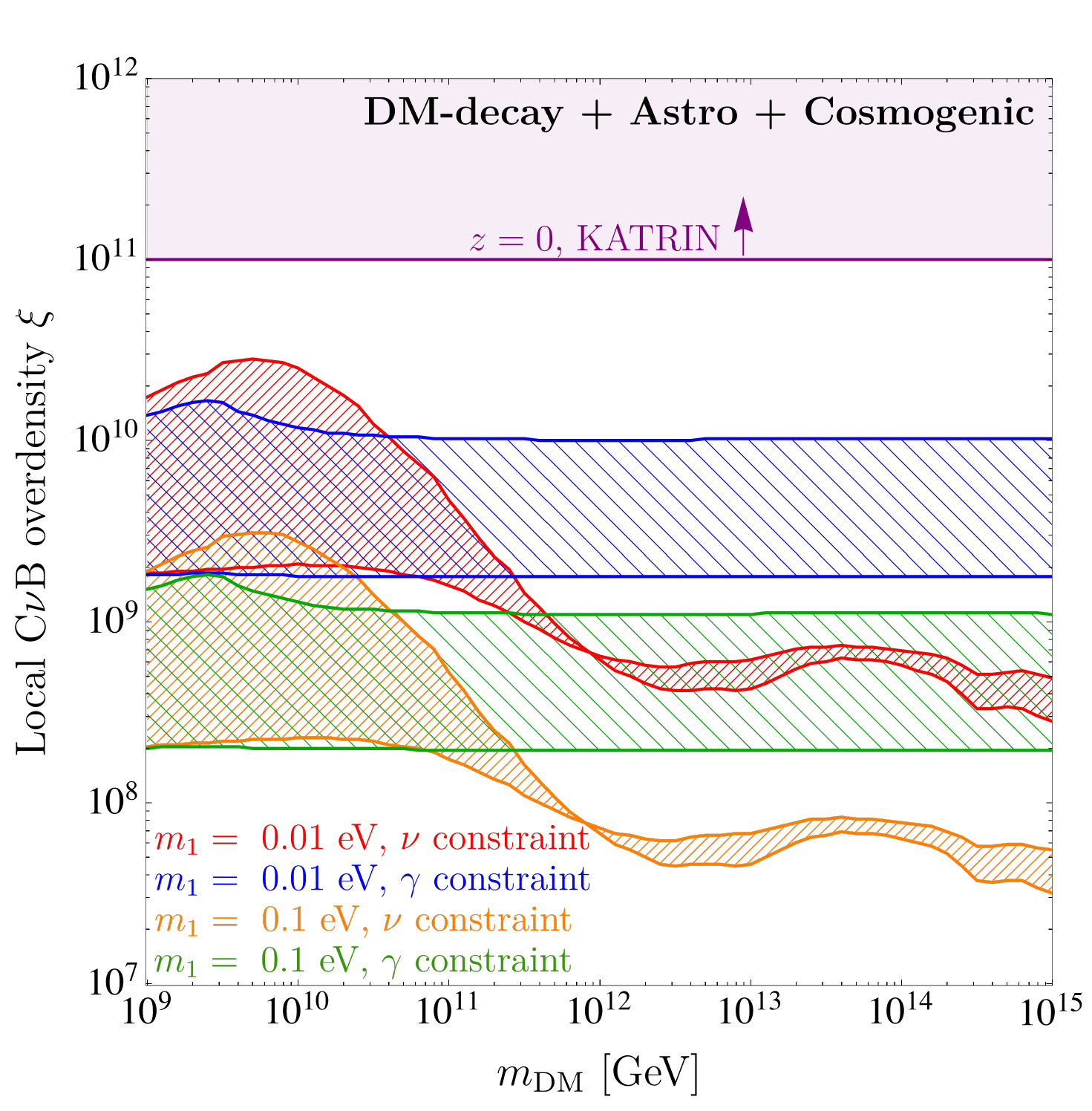}
\caption{\emph{Left:} projected IceCube-Gen2 radio sensitivity at 90\% C.L. on the C$\nu$B overdensity $\xi$ as a function of the lightest neutrino mass, for the case where the neutrino flux includes contributions from both astrophysical sources and DM decay, with the latter computed using both $\gamma$-ray and neutrino constraints. Four benchmark points are shown, corresponding to two values of DM mass and two values of the DM lifetime saturating the constraints.
\emph{Right:}  projected IceCube-Gen2 radio sensitivity at 90\% C.L. on the C$\nu$B overdensity $\xi$ as a function of the DM mass, for the case where the neutrino flux includes contributions from both astrophysical sources and DM decay, with the latter computed using both $\gamma$-ray and neutrino constraints. Four benchmark points are shown, corresponding to two values of neutrino mass and two values of the DM lifetime saturating the constraints.
}
\label{fig:ff_m1}
\end{figure*}

\begin{figure}[t!]
    \centering
    \includegraphics[width=0.99\linewidth]{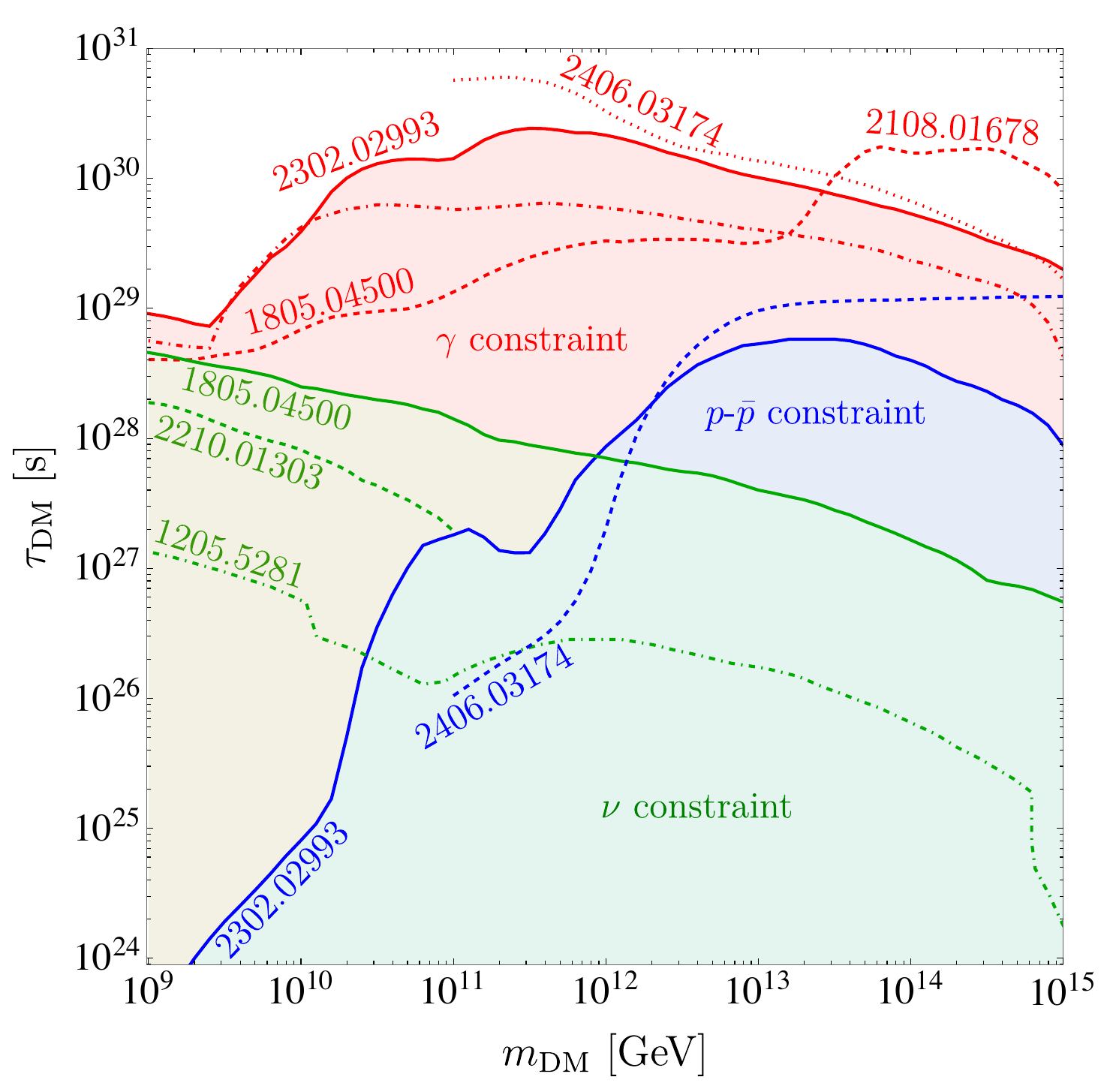}
    \caption{Existing limits in DM mass-lifetime parameter space for neutrinophilic DM. The red shaded region corresponds to the excluded parameter space from the $\gamma$-ray constraints~\cite{Das:2023wtk, Kachelriess:2018rty, Chianese:2021jke, Sarmah:2024ffy} that are typically the strongest. These constraints are derived based on the $\gamma$-ray data collected by experiments like Pierre Auger Observatory (PAO), KASCADE and KASCADE-GRANDE. The blue shaded region denotes the parameter space excluded by the $p-\bar{p}$ constraint which is obtained by fitting the $p-\bar{p}$ data at PAO~\cite{Das:2023wtk, Sarmah:2024ffy}. Finally, the green shaded region represents the excluded parameter space from the nonobservation of UHE neutrinos~\cite{Esmaili:2012us, Kachelriess:2018rty, Arguelles:2022nbl}.}
    \label{fig:tauConstraints}
\end{figure}

Figure~\ref{fig:dm_only_m1} (left panel) shows our projected IceCube-Gen2 radio sensitivity to $\xi$ in the case where only neutrino fluxes from DM decay are considered. The sensitivity is shown as a function of the lightest neutrino mass $m_1$ for two values of DM mass: $m_\mathrm{DM}=10^9\;\mathrm{GeV}$ (red) and $m_\mathrm{DM}=10^{15}\;\mathrm{GeV}$ (orange). 
For the lighter of these, we show two lines corresponding to different DM-induced neutrino fluxes. The solid (dashed) line represents the case where neutrino fluxes are computed such that the $\gamma$-ray (neutrino) DM lifetime constraints in \cref{fig:tauConstraints} are saturated. For $m_\text{DM}=10^9$ GeV these two DM lifetime constraints are comparable; therefore, the expected sensitivity to $\xi$ is of the same order.
The derived sensitivity limits on $\xi$ for $m_\text{DM}=10^9$ GeV are weaker than the KATRIN overdensity bound~\cite{KATRIN:2022kkv} (purple horizontal line), except for $\mathcal{O}$(eV) values of the lightest neutrino mass; that region, however, is in tension with KATRIN's direct mass limit~\cite{KATRIN:2024cdt} (brown vertical line).
For $m_\text{DM}=10^{15}$ GeV, sensitivity down to 
$\xi \gtrsim 10^{6}$ is reachable for the case where the DM-induced neutrino flux is produced such that 
bounds on the DM lifetime from neutrino experiments are saturated. For this value of DM mass, we do not show the scenario with DM-induced neutrino flux that is in accord with DM lifetime limits from $\gamma$-ray experiments, since such neutrino flux would be too small to yield any events at the IceCube Gen-2 radio experiment. For comparison, we also show the cosmological upper limit on the lightest neutrino mass derived from the Planck measurements~\cite{Planck:2018vyg}. It should, however, be noted that the exact value of this limit depends on the cosmological datasets used~\cite{DESI:2025ejh,Chebat:2025kes}. Moreover, it is strictly applicable at high redshifts, and there are ways to relax it at lower redshifts, e.g. by assuming a nonstandard cosmology~\cite{Bellomo:2016xhl,Esteban:2021ozz,Esteban:2022rjk, Craig:2024tky}, in presence of nonstandard neutrino interactions~\cite{Beacom:2004yd,Hannestad:2004qu, Farzan:2015pca,Escudero:2022gez,Benso:2024qrg, Das:2025asx}, decaying neutrinos~\cite{Chacko:2019nej,Escudero:2020ped,FrancoAbellan:2021hdb}, time-varying neutrino mass~\cite{Fardon:2003eh,Dvali:2016uhn,Lorenz:2018fzb,Lorenz:2021alz,deGouvea:2022dtw,Sen:2024pgb}, or neutrinos with a modified distribution function~\cite{Oldengott:2019lke,Alvey:2021sji}. Therefore, we do not shade the region disfavored by standard cosmology.

The right panel of Fig.~\ref{fig:dm_only_m1} exhibits the computed projected upper limit on $\xi$ based on IceCube-Gen2 radio sensitivity as a function of the DM mass $m_\text{DM}$ for four representative values of the lightest neutrino mass: $0.01$ (red), $0.05$ (blue), $0.1$ (orange), and $0.45$ eV (brown).
For each value of the DM mass, the DM lifetime, which controls the magnitude of the induced neutrino fluxes, is taken such that the DM lifetime constraints from neutrino searches are saturated. In this case, we are able to probe overdensities down to $\xi\gtrsim 10^6$. 

Note that generating DM-induced neutrino fluxes saturating the neutrino constraint assumes significant relaxation of the $\gamma$-ray limits. 
The weakening of these limits by an order of magnitude is possible due to uncertainties in the galactic magnetic field modeling~\cite{Munbodh:2024ast}. Another  promising possibility is discussed in Ref.~\cite{Hiroshima:2017hmy}, where DM decays into invisible hidden states in addition to SM neutrinos; this results in relaxed $\gamma$-ray limits as well as in the broadening of the neutrino spectrum.

Now, we consider the scenario in which all available neutrino fluxes--from both astrophysical sources and DM decay--are included in the calculation of the sensitivity limits. This is shown in~\cref{fig:ff_m1}, where in the left panel we present the expected sensitivity for $\xi$ as a function of the lightest neutrino mass $m_1$. The DM-induced part of the total flux is computed for two DM masses ($10^9$ and $10^{15}$ GeV), where for each case the flux is calculated by either saturating neutrino or $\gamma$-ray constraints on the DM lifetime. We observe that both benchmark points computed with $\gamma$-ray constraints on the DM lifetime, as well as the benchmark point $m_\text{DM}=10^{9}$ GeV ($\nu$ constraint), yield similar results, represented by slightly shifted red, green, and blue bands. This signifies that the sensitivity is only mildly impacted by DM, being dominated by cosmogenic and astrophysical neutrinos. The projected sensitivity for C$\nu$B overdensity in this regime varies between $\xi \sim 10^8$ and $\xi \sim 10^{10}$, depending on the value of the lightest neutrino mass. The band illustrates the uncertainty in the theoretical modeling of the cosmogenic fluxes (see, e.g., Ref.~\cite{Muzio:2023skc}). In the left panel of ~\cref{fig:ff_m1}, we also observe that a significant relaxation of the $\gamma$-ray constraints could further improve these predictions, as the DM-induced neutrino flux would not only become competitive with but even surpass the astrophysical and cosmogenic flux. This occurs for large DM masses, as represented by the orange band calculated using a DM mass of $10^{15}$ GeV and a lifetime taken according to neutrino constraints. For this case, we find a projected sensitivity on $\xi$ reaching down to approximately $10^6$, thereby converging to the respective DM-only case shown in the left panel of~\cref{fig:dm_only_m1}.

This effect can also be seen in the right panel of~\cref{fig:ff_m1}, where for $m_\mathrm{DM} \gtrsim 10^{11}$ GeV the sensitivity limits on the local C$\nu$B overdensity calculated using the saturated neutrino DM lifetime constraints (densely hatched region) start to dominate over the limits calculated using the saturated $\gamma$-ray DM lifetime constraints (sparsely hatched region). Note that this effect is present irrespective of the value of the lightest neutrino mass. In this regime, where the DM-induced neutrino flux dominates over the cosmogenic flux, there is less impact of the astrophysical uncertainty due to UHECR modeling on the overdensity sensitivity.

Once the UHE neutrino flux is observed, distinguishing whether it corresponds to an attenuated spectrum due to a local C$\nu$B overdensity or an unattenuated flux with a different normalization is crucial. 
We outline several ways to approach this problem. First, the UHE neutrino fluxes in these two cases would exhibit different neutrino energy dependencies, which future precise spectral measurements could disentangle~\cite{IceCube-Gen2:2020qha}. Second, independent $\gamma$-ray observations could reveal the presence of a DM component, thereby favoring an overdensity interpretation. Third, future results from direct probes of local neutrino overdensities in experiments such as KATRIN~\cite{KATRIN:2022kkv}, Project-8~\cite{Project8:2014ivu}, HOLMES~\cite{Alpert:2025tqq}, or PTOLEMY~\cite{PTOLEMY:2018jst} may provide complementary evidence. Last but not least, improved modeling of the cosmogenic neutrino flux, driven by a more precise determination of the UHE cosmic ray composition, would reduce theoretical uncertainties and help discriminate between the two cases.

\section{Summary and Conclusions}
\label{sec:Conclusion}

The decay of heavy neutrinophilic DM can produce UHE neutrinos that may scatter off the C$\nu$B. Astrophysical and cosmogenic neutrinos can also undergo such scattering. In the presence of a large local C$\nu$B overdensity, such interactions can significantly modify the UHE neutrino flux, making it possible to test for these overdensities.

In this work, we have calculated projected constraints on the local C$\nu$B overdensity using the future observations of the UHE neutrinos in the next-generation experiment IceCube-Gen2 radio. We considered two scenarios: (i) the UHE neutrino flux originates solely from DM decay, and (ii) the flux includes contributions from DM decay, as well as astrophysical and cosmogenic neutrinos.
In case (i), if the neutrino flux is computed in a way that respects $\gamma$-ray limits on DM decay, the projected sensitivity to the C$\nu$B overdensity is generally weaker than current limits, except for large neutrino masses. However, the situation improves significantly, especially for DM masses above approximately $10^{11}$ GeV, if those $\gamma$-ray constraints are relaxed, since a larger DM-induced neutrino flux becomes allowed. In such case, we find that the projected sensitivity to the local C$\nu$B overdensity reaches as low as $\sim 10^6$. 
In case (ii), where we considered UHE neutrinos produced from both DM decays and in astrophysical settings, we generally find the projected sensitivity between $\xi\sim 10^8$ and $\xi \sim 10^{10}$, depending on the value of the lightest neutrino mass, if the $\gamma$-ray constraints are respected. The sensitivity further increases and converges to scenario (i) when neutrino fluxes from heavy DM are considered in a scenario with relaxed $\gamma$-ray constraints. Overall, our results point to an interesting novel direction for probing the C$\nu$B clustering in the near future. This is complementary to other direct and indirect searches for the C$\nu$B. At the same time, it could shed some light on the nature of dark matter. 

\acknowledgments
\noindent
We would like to thank Carlos Arg\"{u}elles, Jose Carpio, Alex Chen, Bryce Cyr, Saikat Das, Kohta Murase, and Jack Shergold  for useful discussions. The work of VB is supported by the U.S. Department of Energy Grant No. DE-SC0025477. The work of WM and BD was partly supported by the U.S. Department of Energy under Grant No.~DE-SC0017987. BD was also supported in part by a Humboldt Fellowship from the Alexander von Humboldt Foundation. The work of AMS was supported in part by Department of Energy Grant No.\ DE-AC02-07CH11359: \emph{Neutrino Theory Network Program}. We would like to thank the Center for Theoretical Underground Physics and Related Areas (CETUP*) and the Institute for Underground Science at Sanford Underground Research Facility (SURF) for hospitality and for providing a stimulating environment during the 2023, 2024 and 2025 summer workshops, where parts of this work were conducted. We would also like to thank the organizers of PHENO 2022 at University of Pittsburgh, where the initial idea of this project was generated.  

\appendix

\section{DM decay constraints}
\label{sec:DMtau}
In \cref{fig:tauConstraints}, we summarize the existing limits in the heavy DM mass--lifetime parameter space, coming from $\gamma$-ray~\cite{Das:2023wtk, Kachelriess:2018rty, Chianese:2021jke, Sarmah:2024ffy}, $p$-$\bar{p}$~\cite{Das:2023wtk,Sarmah:2024ffy}, and neutrino~\cite{Esmaili:2012us, Kachelriess:2018rty, Arguelles:2022nbl} observations. In all these cases, we have assumed that the DM primarily decays into neutrinos. However, other final states appear unavoidably as secondaries through the electroweak bremsstrahlung. These limits are used to set the benchmarks studied in the main text. 

\section{SHDM production}\label{sec:DMprod}
There are various nonthermal production mechanisms for a SHDM that can yield the correct relic density. Here we briefly mention two commonly discussed mechanisms: 
\subsection{Gravitational production}
This is the most model-independent mechanism, which only
requires the DM coupling to gravity.  SHDM is produced by  vacuum fluctuations in the expanding spacetime background during or at the end of inflation~\cite{Chung:1998zb, Kuzmin:1998kk}, with a relic density today of  order~\cite{Chung:2001cb} 
\begin{align}
    \Omega_{\rm DM}h^2\sim \left(\frac{m_{\rm DM}}{10^{11}~{\rm GeV}}\right)^2\left(\frac{T_{\rm RH}}{10^{9}~{\rm GeV}}\right) \, .
    \label{eq:grav}
\end{align}
Thus, for a given SHDM mass, an appropriate reheating temperature $T_{\rm RH}$ can be chosen to yield the observed relic density of $\Omega_{\rm DM}h^2=0.12$~\cite{Planck:2018vyg}. The exact numerical coefficients in Eq.~\eqref{eq:grav} depend on the model of inflation (e.g., chaotic versus hybrid), but this scaling relation is robust, so long as $m_{\rm DM}<H_I$, the Hubble expansion rate during inflation. For $m_{\rm DM}>H_I$, the particle production is exponentially suppressed and the result will depend on whether the particle was relativistic or nonrelativistic at the end of inflation~\cite{Chung:1998bt}. 

\subsection{Inflaton decay}
If the inflaton $\phi$ couples to the SHDM $X$, its decay $\phi \to XX$ can also produce a nonthermal population of DM~\cite{Allahverdi:2002nb, Allahverdi:2002pu, Dev:2013oiy, Bernal:2021qrl, Bernal:2024ykj}, with the current relic density given by
\begin{align}
    \Omega_{\rm DM}h^2\simeq & 0.12\left(\frac{m_{\rm DM}}{10^{11}~{\rm GeV}}\right)\left(\frac{T_{\rm RH}}{10^{9}~{\rm GeV}}\right)\left(\frac{10^{13}~{\rm GeV}}{m_\phi}\right)\nonumber \\
    &\qquad \times \left(\frac{{\rm Br}}{10^{-8}}\right) \, .
    \label{eq:infl}
\end{align}
Thus, a tiny decay branching ratio (Br) of $\phi\to XX$ is enough to generate the observed DM relic density. Note that Eq.~\eqref{eq:infl} assumes perturbative inflaton decay and negligible annihilation or thermalization of 
DM after production. For the modified expression when backreaction, scatterings, or nonperturbative preheating are important, see  Ref.~\cite{Bernal:2024ykj}.

This mechanism can be generalized to the decay of any matterlike component that drives an early-matter-domination phase. A well-known example is moduli decay in supersymmetric theories~\cite{Moroi:1999zb,Gelmini:2006pw}. In string theory, generic features of string compactifications, such as high-scale supersymmetry breaking that gives rise to successful inflation and epochs of early matter domination driven by string moduli, also provide a suitable framework to obtain the correct relic density for SHDM~\cite{Allahverdi:2013noa,Allahverdi:2020uax, Allahverdi:2023nov}.

\section{Neutrino transport equation}
\label{sec:transport}

The neutrino transport equation captures the cosmological evolution of the comoving number of neutrinos per unit energy at time $t$, $\tilde{N}_A(E,t) = dN_A/dE$ for a given neutrino species $A$. The neutrino source located at a particular redshift $z_s$ injects the neutrinos at time $t=0$. The comoving number of neutrinos per unit energy at the redshift $z_s$ is obtained using \texttt{HDMSpectra}~\cite{Bauer:2020jay}. These neutrinos are then propagated until the flux reaches Earth. 
The complete neutrino transport equation reads~\cite{Das:2024bed}
\begin{align}
\frac{\partial \tilde{N}_{A}(E,t)}{\partial t}& = \frac{\partial}{\partial E} [H(z(t))E\tilde{N}_{A}(E,t)]  -\tilde{N}_{A}(E,t)\mathcal{A}_{A}(E,t)  \nonumber \\
& + \int_E^\infty dE' \tilde{N}_{A}(E',t) \mathcal{B}_{A\to A}(E,E')\nonumber \\
& + \sum_{B\neq A}\int_E^\infty dE'\tilde{N}_{B}(E',t)\mathcal{C}_{B\to A}(E,E')\nonumber \\
& +\tilde{\mathcal{Q}}_{A}(E,t) \ ,
\label{TransportEquation_Time}
\end{align}
where $A\in \{\nu_1,\nu_2,\nu_3,\bar{\nu}_1,\bar{\nu}_2,\bar{\nu}_3\}$ are the neutrino mass eigenstates, which are connected to the flavor eigenstates of neutrinos $\nu_{\alpha}$ by the Pontecorvo–Maki–Nakagawa–Sakata (PMNS)  mixing matrix, assumed to be unitary. The first term on the right-hand side of Eq.~\eqref{TransportEquation_Time} is the adiabatic loss term due to the expansion of the Universe. The last term  $\tilde{\mathcal{Q}}_A(E,t)$ is the source term, accounting for the number of neutrinos of species $A$ with energy $E$ injected by the source per unit time per unit energy at a particular time $t$. We assume $\tilde{\mathcal{Q}}_A(E,t) = \delta(t){dN_A}/{dE}$. The $\mathcal {A}_A(E,t)$-term in Eq.~\eqref{TransportEquation_Time} is the total absorption rate of neutrino species $A$ with energy $E$, $\mathcal{B}_{A\to A}$ is the self-production rate of species $A$ with energy $E$ being created from the same particle type $A$ but with different energy $E'$, and $\mathcal{C}_{B\to A}$ is the production rate of species $A$ with energy $E$ from other neutrino species $B\neq A$ with energy $E'$. Both $\mathcal{B}$ and $\mathcal{C}$ terms provide the reinjection of neutrinos into the flux. For details, see Ref.~\cite{Das:2024bed}.

\bibliography{refs}

\end{document}